\newcommand{\Tr}{\mathop{\mathrm{Tr}}\nolimits}

\documentclass[pra,twocolumn,showpacs,superscriptaddress,eqsecnum,10pt]{revtex4-1}

\usepackage{amsmath,amssymb,graphicx,color,mathptmx,hyperref,bm}
 
\begin{document}

\title{Classical distinguishability as an operational measure of
  polarization}

\author{G.~Bj\"{o}rk} 
 \affiliation{Department of Applied Physics,
  Royal Institute of Technology (KTH), AlbaNova Center, SE-106 91
  Stockholm, Sweden}

\author{H.~de Guise} 
\affiliation{Department of Physics, Lakehead University,
  Thunder Bay, ON~P7B 5E1, Canada}

\author{A.~B.~Klimov} 
\affiliation{Departamento de F\'{\i}sica,
  Universidad de Guadalajara, 44420~Guadalajara, Jalisco, Mexico}

\author{P.~de la Hoz} 
\affiliation{Departamento de \'Optica, Facultad
  de F\'{\i}sica, Universidad Complutense, 28040~Madrid, Spain}

\author{L.~L.~S\'{a}nchez-Soto} 
\affiliation{Departamento de \'Optica,
  Facultad de F\'{\i}sica, Universidad Complutense, 28040~Madrid,
  Spain}

\begin{abstract}
  We put forward an operational degree of polarization that can be
  extended in a natural way to fields whose wave fronts are not
  necessarily planar.  This measure appears as a distance from a state
  to the set of all its polarization-transformed counterparts. By
  using the Hilbert-Schmidt metric, the resulting degree is a sum of
  two terms: one is the purity of the state and the other can be
  interpreted as a classical distinguishability, which can be
  experimentally determined in an interferometric setup.  For
  transverse fields, this reduces to the standard approach, whereas it
  allows one to get a straight expression for nonparaxial fields.
\end{abstract}


\maketitle

\section{Introduction}

Far from its source, any electromagnetic wave can be locally
approximated by a plane wave; i.e., with a well-defined direction of
propagation and thus a specific transverse plane.  Such beamlike
fields are described by two orthogonal electric-field components and,
consequently, their polarization is characterized by a $2 \times
2$ correlation matrix, usually called the polarization
matrix~\cite{Brosseau:1998lr,Mandel:1995qy}.

This polarization matrix can be uniquely decomposed as a sum of two
matrices: one represents a fully polarized part and the other
a completely unpolarized part. The ratio of the intensity of the
polarized part to the total intensity is the degree of polarization.

Equivalently, one may resort to the Stokes parameters, which are the
coefficients of the expansion of the polarization matrix onto the
Pauli basis.  These variables determine a locus on the Poincar\'e
sphere, wherein the state of polarization is elegantly visualized:
actually, the degree of polarization can be seen as the length of the
Stokes vector.

This two-dimensional (2D) theory is the backbone of the standard
polarization optics. However, the necessity of addressing new issues,
such as highly nonparaxial fields~\cite{Nicholls:1972zr}, narrow-band
imaging systems~\cite{Pohl:1984fk}, and the recognition of associated
propagation questions~\cite{Petruccelli:2010ly}, has revived
interest in extending the 2D approach to fully three-dimensional
(3D) field distributions. Although this question has been
considered for many years, no satisfactory solution has thus far been
found.  Indeed, there are several contradictory claims made in the
literature on this subject~\cite{Samson:1973ve,Barakat:1977qf,
 Setala:2002gb,Setala:2002oq,Korotkova:2004rv,Luis:2005kl,
 Ellis:2005bh,Refregier:2006fj,Dennis:2007tg,Sheppard:2011hc}.  The
divergences occur because notions that are equivalent for the 2D
case, lead to different definitions when extrapolated to the 3D limit.
This diversity has prompted various authors to suggest alternative 3D
measures of polarization based in, e.g., non-quantum
entanglement~\cite{Qian:2011kx}, von Neumann
entropy~\cite{Gil:2007fk}, the fully polarized field
component~\cite{Ellis:2005bh}, or the invariants of the rotational
group~\cite{Barakat:1983xy}. All of these instances produce sensible
computable magnitudes, but they are hardly measurable, which prevents 
a proper assessment of their merits.

In this paper, we revisit an operational measure introduced
some time ago in the realm of quantum optics~\cite{Bjork:2002br}: in
2D, the prescription is to look at the minimum overlap
between a state and the set of its polarization-transformed [i.e.,
SU(2)-rotated] counterparts.  The key point is that this magnitude, as
discussed in Ref.~\cite{Bjork:2000kx}, can be directly determined as
the visibility of an interference experiment.  Our main goal is to
extend this notion to the 3D case.

To this end, we first reinterpret that measure as a distance between
the state and its rotated partners. In this vein, it is worth
stressing that distance measures have been successfully employed in
assessing a number of disputed quantities, such as
nonclassicality~\cite{Hillery:1987kq,Dodonov:2000lq,Marian:2002ao},
entanglement~\cite{Vedral:1997cj,Marian:2008um,Bellomo:2012qf},
information~\cite{Gilchrist:2005wu,Ma:2009jx,Monras:2010il},
non-Gaussianity~\cite{Genoni:2007lk}, and
localization~\cite{Mirbach:1998ti,Gnutzmann:2001fs}, to cite only a
few examples.

Two main hurdles are usually faced when defining a distance-type
measure: choosing a convenient metric and identifying a reference set
of states.  As to the first question, different candidates have been
investigated, including among others, relative
entropy~\cite{Wehrl:1978ug,Vedral:2002zh,Ohya:2004it}, Bures and
related metrics~\cite{Bures:1969kg,Uhlmann:1976jf,Wootters:1981yp,
  Braunstein:1994ni}, as well as Monge~\cite{Zyczkowski:1998gf},
trace~\cite{Belavkin:2005jf,Nielsen:2010ft}, and
Hilbert-Schmidt~\cite{Witte:1999ta,Ozawa:2000uv, Bertlmann:2002vb}
distances, each having its own advantages for certain applications. In
particular, the last one is probably the simplest from a computational
viewpoint and will be adopted here.

In polarization, it has been suggested to take unpolarized states as
the reference set, both in the quantum~\cite{Klimov:2005kl} and the
classical domain~\cite{Luis:2007in}. Such a set is very well
characterized~\cite{Prakash:1971fr,Agarwal:1971zr,Lehner:1996rr} and
this provides sensible results. However, as anticipated above, we
prefer to consider the rotated versions of the original state. Going
from the 2D to the 3D situation is just extending the SU(2)-rotated
set to its SU(3) analog, and the resulting degrees have a clear
physical interpretation.

The paper is arranged as follows. In Sec.~\ref{sec:dop} we recall the
basic tools used to describe the partial polarization of both 2D and
3D electromagnetic fields, emphasizing the similarities and
differences between these two situations. In Sec.~\ref{sec:distbas},
we introduce the general notion of degree of polarization as a
distance, and work out the resulting expressions for both cases,
comparing with previous proposed measures. Finally, we summarize our
work in Sec.~\ref{sec:conc}.

\section{Basic description of polarization}
\label{sec:dop}

A pivotal quantity in the characterization of polarization
of both 2D and 3D fields is the degree of polarization. It
quantitatively captures the random character of the electric field as a
function of time. Such a behavior cannot be accounted for in terms of
a deterministic description: we must, instead, adopt a statistical
perspective. To be as self-contained as possible, we briefly review
the essential ingredients needed for that purpose.

\subsection{Two-dimensional fields}

Consider a  monochromatic beam propagating in the $z$
direction. The electric field can be resolved in the transverse plane
in terms of horizontal ($x$) and vertical ($y$) components, which are
taken to be a probabilistic ensemble given by $E_{x}$ and $E_{y}$. The
corresponding $2 \times 2$ (equal-time) polarization matrix (also
called the coherence matrix) is defined
as~\cite{Brosseau:1998lr,Mandel:1995qy}
\begin{equation}
  \varrho_{k \ell}^{(2)} =  
  \langle E_{k}^{\ast} E_{\ell} \rangle \, , 
  \qquad
  k, \ell \in \{ x, y\} \, .
\end{equation}
Here, the brackets denote ensemble averaging over different
realizations and the superscript indicates the dimensionality, although
in the following we will suppress it when there is no risk of
confusion. 

The diagonal elements of the matrix $\varrho$ represent the energy
distribution between the two components of the field: $ I = \langle |
E_{x} |^{2} \rangle + \langle | E_{y} |^{2} \rangle = \Tr (\varrho )$,
where $\Tr$ is the trace of the matrix. Without loss of generality, we
henceforth normalize this intensity to unity.  On the other
hand, the off-diagonal elements describe the correlations between the
field components. From its very definition, it follows that
$\varrho_{k\ell} = \varrho_{\ell k}^{\ast}$, so $\varrho$ is
Hermitian.

The matrix $\varrho$ can be conveniently decomposed in terms of the
(Hermitian) Pauli matrices $\bm{\sigma}$; the result reads
\begin{equation}
  \label{eq:genP}
  \varrho =\frac{1}{2} 
  \left ( \openone + \mathbf{n} \cdot \bm{\sigma} \right )  \, .
\end{equation}
The normalized coordinates $n_{r}$ ($r=1,2,3$) can be recovered as
\begin{equation}
  \label{eq:defStok}
  n_{r} = \Tr ( \varrho \, \sigma_{r} ) \, ,
\end{equation}
and are nothing but the Stokes parameters. In other words, we can
map each polarization matrix $\varrho$ into a Stokes vector $\varrho
\mapsto \mathbf{n} = (n_{1}, n_{2}, n_{3})$.  The length of
$\mathbf{n}$ will be denoted as
\begin{equation}
  \label{eq:defPle}
  \mathbb{P}^{(2)}  = | \mathbf{n} | = 
  \sqrt{n_{1}^{2} + n_{2}^{2}  + n_{3}^{2}} \, ,
\end{equation}
and, as we shall justify soon, deserves the name of degree of
polarization for 2D fields.

The Stokes parameters provide geometric information about the
polarization ellipse; i.e., the ellipse that the electric field tip traces
out during one optical cycle.  The parameters $n_{1}$ and $n_{2}$
carry information about the alignment of the ellipse axes, while $ \pi
n_{3}$ gives the ellipse area, signed according to polarization
handedness.

If the relation between the $E_{x}$ and $E_{y}$  is completely
deterministic, the field is fully polarized. For such a pure state
(borrowing the terminology from quantum optics), the polarization
matrix is idempotent, i.e.,
\begin{equation}
  \varrho_{\mathrm{pol}}^{2} = \varrho_{\mathrm{pol}} \, ,
\end{equation}
and we get $\mathbb{P}^{(2)}_{\mathrm{pol}} = 1$. On
the other hand, if the components of the field are fully uncorrelated,
the off-diagonal elements are zero. If, in addition, the energy is
distributed evenly between the $x$ and $y$ components, 
\begin{equation}
  \varrho_{\mathrm{unpol}} = 
  \textstyle{\frac{1}{2}} \, \openone \, , 
\end{equation}
and we have $\mathbb{P}^{(2)}_{\mathrm{unpol}} = 0$. This leads to the
important decomposition of $\varrho$ into fully polarized and
unpolarized parts, viz.,
\begin{equation}
  \label{eq:decpu}
  \varrho = [ 1 - \mathbb{P}^{(2)} ]  \varrho_{\mathrm{unpol}} + 
  \mathbb{P}^{(2)}   \varrho_{\mathrm{pol}} \, .
\end{equation}
In this way, $\mathbb{P}^{(2)}$ appears as the proportion of the
energy of the fully polarized part from the total energy, which gives
a transparent physical meaning to the definition of $\mathbb{P}^{(2)}$.

Alternatively, $\mathbb{P}^{(2)}$ can be written in a slightly
different yet equivalent way,
\begin{equation}
  \label{eq:defPur}
  \mathbb{P}^{(2)}  = \sqrt{2 \Tr (\varrho^{2} ) - 1} =  
  \sqrt{1 - 4 \det (\varrho})    \, ,
\end{equation}
as can be checked by a direct calculation. In the first form, the
degree of polarization seems to be intimately linked to $\Tr
(\varrho^{2} )$, which, following again a quantum jargon, is called
the  purity. In the second form, it can be immediately related with the
eigenvalues of $\varrho$: if we denote them by $\lambda_{+}$ and
$\lambda_{-}$,  ($\lambda_{+} > \lambda_{-}$),  then $(\lambda_{+} +
\lambda_{-})^{2} = 1$ and $\det  (\varrho) = \lambda_{+} \lambda_{-}$,
so that 
\begin{equation}
  \label{eq:defPeig}
  \mathbb{P}^{(2)}   = \lambda_{+} - \lambda_{-} \, ,
\end{equation}
the importance of which will soon become apparent.

Polarization transformations are generated by wave plates and
represented by $2\times 2$ unitary matrices of
SU(2)~\cite{Simon:1989ly},
\begin{eqnarray}
  \label{eq:su2mat}
  R_{g} & \equiv & R(\alpha, \beta, \gamma) \nonumber \\
  & = &  
  \left(
    \begin{array}{cc}
      e^{- i (\alpha +\gamma )/2} \cos (\beta /2) & 
      -e^{-i (\alpha -\gamma )/2} \sin ( \beta /2 ) \\
      e^{+ i (\alpha -\gamma )/2} \sin (\beta /2 ) & 
      e^{+ i (\alpha +\gamma )/2} \cos ( \beta /2 ) \\
    \end{array}
  \right) \, , \nonumber \\
\end{eqnarray}
where $(\alpha, \beta, \gamma)$ denote the Euler angles. The action of
these transformations on the polarization matrix is via conjugation,
\begin{equation}
  \label{eq:conj}
  \varrho_{g} = R_{g} \, \varrho R_{g}^{\dagger} \, ,
\end{equation}
which, in turn, induces rotations on the Stokes vector $\mathbf{n}$,
as confirmed by the well-known relation between SU(2) and the group of
rotations SO(3) ~\cite{Cornwell:1997bh}. The essential point is that $
\mathbb{P}^{(2)} $ is clearly unchanged by these transformations.

\subsection{Three-dimensional fields}

Next, we loosen the restriction to planar geometry and examine the
behavior of electric fields having three nonvanishing components, in
directions we denote as $x$, $y$, and $z$, respectively. Now, the
vibrations of the field are not constrained to a plane and the
polarization must be described by a $3 \times 3$ matrix
\begin{equation}
  \label{eq:pol3d}
  \varrho^{(3)}_{k \ell} = \langle E_{k} E_{\ell} \rangle \, ,
  \qquad k, \ell \in \{x, y, z \} \, .
\end{equation}
The superscript 3 labels the 3D approach and will be dropped when
the context is clear.

If all of the components are completely uncorrelated (and their
energies are equal) the field is unpolarized and its direction 
is random.  If one of the components has less energy than the
other two, the vibrations are less random and, consequently, the field
is more polarized than in the equal-energy case. This means that any
field having only two non-vanishing components is never unpolarized in
the 3D sense, regardless of the correlations between the
components. Hence, a planar field, which is commonly called
unpolarized in 2D, is not fully unpolarized but partially polarized
in a 3D description.

As in 2D, the field is called fully polarized if all of the
field components are completely correlated. Hence, in contrast to an
unpolarized field, a planar field that is fully polarized is always
fully polarized also in the 3D sense.

One of the most remarkable differences between 2D and 3D is that the
$3 \times 3$ polarization matrix cannot be generally expressed as a
sum of unpolarized and fully polarized
parts~\cite{Ellis:2005bh}. Therefore, if one desires to define a
degree of polarization for arbitrary electric fields, the approach
taken in \eqref{eq:decpu} must be abandoned.

In any event, the $3\times 3$ polarization matrix can be expanded in a
basis as
\begin{eqnarray}
  \rho & = & \frac{1}{3} \left ( \openone + \sqrt{3} \,
    \mathbf{n}  \cdot \bm{\Lambda} \right ) \, ,
\end{eqnarray}
where $\bm{\Lambda}$ are the Gell-Mann matrices (see details
in the Appendix).  The corresponding coordinates of the
eight-dimensional Stokes vector can be obtained as
\begin{equation}
  \label{eq:3}
  n_{r} = \frac{\sqrt{3}}{2} \Tr (\varrho \Lambda_{r}) \, .
\end{equation}
We have introduced the factor $\sqrt{3}$ in such a way that
for a pure state $\mathbf{n} \cdot \mathbf{n} =
1$~\cite{Arvind:1997qf}, although other choices can be found in the
literature.  One first option would be to define~\cite{Setala:2002oq}
\begin{equation}
  \label{eq:defPle3}
  \mathbb{P}^{(3)}  = | \mathbf{n} | = 
  \sqrt{\sum_{r=1}^{8} n_{r}^{2} } \, ,
\end{equation}
i.e., again the length of the Stokes vector, which is readily shown to
verify $0 \le \mathbb{P}^{(3)} \le 1$. Although this is mathematically
correct, it is not clear physically what $\mathbb{P}^{(3)}$
represents.  Unlike in 2D, where the Stokes vector represents
the complete state of polarization and can be easily visualized, the
generalized Stokes vector is eight dimensional and the geometrical
space supporting this vector is not intuitive at all.

An alternative is to generalize \eqref{eq:defPur} in a way so as to
get the appropriate normalization; it
reads~\cite{Samson:1973ve,Barakat:1977qf}
\begin{equation}
  \label{eq:defPur3}
  \mathbb{P}^{(3)}  = \sqrt{\frac{3 \Tr (\varrho^{2} ) - 1}{2}}  \, .
\end{equation}
The drawback of this definition is that it cannot be understood as a
portion of the energy of the fully polarized part from the total
energy and hence its physical properties need further examination.

Finally, the generalization of \eqref{eq:defPeig} seems even more
dubious, since now we have three different eigenvalues. This
reveals the major problem when extending 2D to 3D instances:
while one parameter is enough to specify the degree of polarization in
2D, two independent parameters are, in general, needed when
considering 3D, which makes the transition a tricky business.

We complete this section by describing the polarization
transformations possible in the 3D case: they are represented by
$3\times 3$ matrices of SU(3), which we write as~\cite{Rowe:1999dn}
\begin{align}
  \label{su3}
  R_{g} & = R_{g} (\Omega) \equiv T_{23}(\alpha_1,\beta_1,-\alpha_1)
  T_{12}(\alpha_{2},\beta_2,-\alpha_2) \nonumber\\
  & \times T_{23}(\alpha_3,\beta_3,-\alpha_3) \Phi(\gamma_1,\gamma_2)
\end{align}
where $\Omega$ is an octuple of Euler-like angles $ \Omega=
(\alpha_1,\beta_1,\alpha_2,\beta_2,\alpha_3,\beta_3,\gamma_1,\gamma_2
)$ and the set $\{T_{ij}\}$ comprises SU(2) subgroup matrices
\begin{equation}
  \label{eq:r23}
  T_{23}   =
\begin{pmatrix} 
    1 & 0 &  0 \\
    0 & e^{- i (\alpha+\gamma )/2}  \cos(\beta/2)
    & - e^{-i (\alpha -\gamma )/2}\sin (\beta /2 ) \\
    0 &  e^{+ i(\alpha -\gamma )/2} \sin (\beta /2) & 
    e^{+ i(\alpha +\gamma )/2} \cos (\beta / 2) 
  \end{pmatrix} ,
\end{equation}
or
\begin{equation}
  T_{12}   =
\begin{pmatrix}
    e^{-i (\alpha+\gamma )/2}\cos (\beta/2 )
    &- e^{- i (\alpha-\gamma)/2} \sin (\beta /2 )  & 0 \\ 
    e^{+ i (\alpha -\gamma )/2} \sin (\beta /2)  & 
    e^{+ i (\alpha +\gamma )/2} \cos (\beta/2 ) &0 \\
    0 & 0 & 1
  \end{pmatrix},
  \label{eq:r12}
\end{equation}
depending on the values of $(ij)$. Also,
\begin{equation}
  \Phi(\gamma_1,\gamma_2)
  =\text{diag} ( e^{-2i \gamma_1}, e^{i(\gamma_1-\gamma_2/2)},
  e^{i (\gamma_1+\gamma_2/2)}).
  \label{diagonalh}
\end{equation}
Equation~(\ref{su3}) factorizes then into SU(2) submatrices, 
with parameters defined by the corresponding Euler angles.  

The action of these transformations on $\varrho$ is via
conjugation as in \eqref{eq:conj}, which induces rotations on the
vector $\mathbf{n}$. However, one word of caution seems pertinent
here: there is no obvious physical interpretation via optical elements
of SU(3) transformations, as now the plane waves averaging to the $3
\times 3$ polarization matrix do not share a common propagation
direction, in general.  Any physical device represented by a SU(3)
transformation should be insensitive to the propagation directions of
the separate members of the ensemble~\cite{Dennis:2004xq}.

Despite the recent progress achieved in the control and manipulation
of 3D polarization~\cite{Li:2012xe}, we are still far from having at
our disposal an SU(3) gadget, in sharp contrast with the simplicity of
SU(2).  Given these experimental difficulties, one might be tempted to
consider invariance only under rotations and inversions; that is, a
field is less polarized at a point if its behavior is fairly unchanged
after we rotate it and reflect it around that point. Although
attractive, this proposal does not allow us to find analytical results
in what follows. Accordingly, we take SU(3) as the symmetry of the
problem, even if its operational implementation may be elusive.

\section{Operational degree of polarization}
\label{sec:distbas}

As heralded in the Introduction, our proposal for the degree of
polarization starts from the \textit{ansatz}
\begin{equation}
  \label{eq:defsw} 
  \mathbb{P}^{(n)} (\varrho) \propto \sup_{g \in \mathrm{SU}(n)}  \,  
  D ( \varrho , R_g  \varrho R_g^\dagger ) \, .
\end{equation}
Here, the $\sup_{g}$ is taken over SU(2) or SU(3), depending on the
appropriate situation. In addition, $D(\varrho,
\varrho^{\prime} )$ stands for any measure of distance between the
polarization matrices $\varrho$ and $\varrho^{\prime}$.

It is clear that there are numerous nontrivial choices for $D(\varrho
, \varrho^{\prime} )$ (by nontrivial we mean that the choice is not a
simple scale transformation of any other distance). None of them could
be said to be more important than any other \textit{a priori}, but the
significance of each candidate would have to be seen through physical
assumptions.  In our case, we take the Hilbert-Schmidt distance
\begin{eqnarray}
  D_{\mathrm{HS}}^2 (\varrho, \varrho^{\prime} ) & = &
  \frac{1}{2} \Tr [ (\varrho - \varrho^{\prime})^2 ] \nonumber \\
  & = & 
  \frac{1}{2} [  \Tr ( \varrho^{2})  + \Tr (\varrho^{\prime 2} ) - 
  2 \Tr( \varrho \varrho^{\prime} )] \, .
\end{eqnarray}
Since $\varrho^{\prime} = \varrho_g = 
R_g\varrho R_g^\dagger$ and $ \Tr (\varrho^{2}) = 
 \Tr (\varrho_{g}^{2} )$,  this distance reduces to
\begin{equation}
  \label{eq:HSdsim}
  D_{\mathrm{HS}}^2 (\varrho, \varrho_{g} ) =
  \Tr(\rho^2) - \Tr  (\varrho \varrho_g) \, .
\end{equation}

\begin{figure}[b]
  \includegraphics[width=0.90\columnwidth]{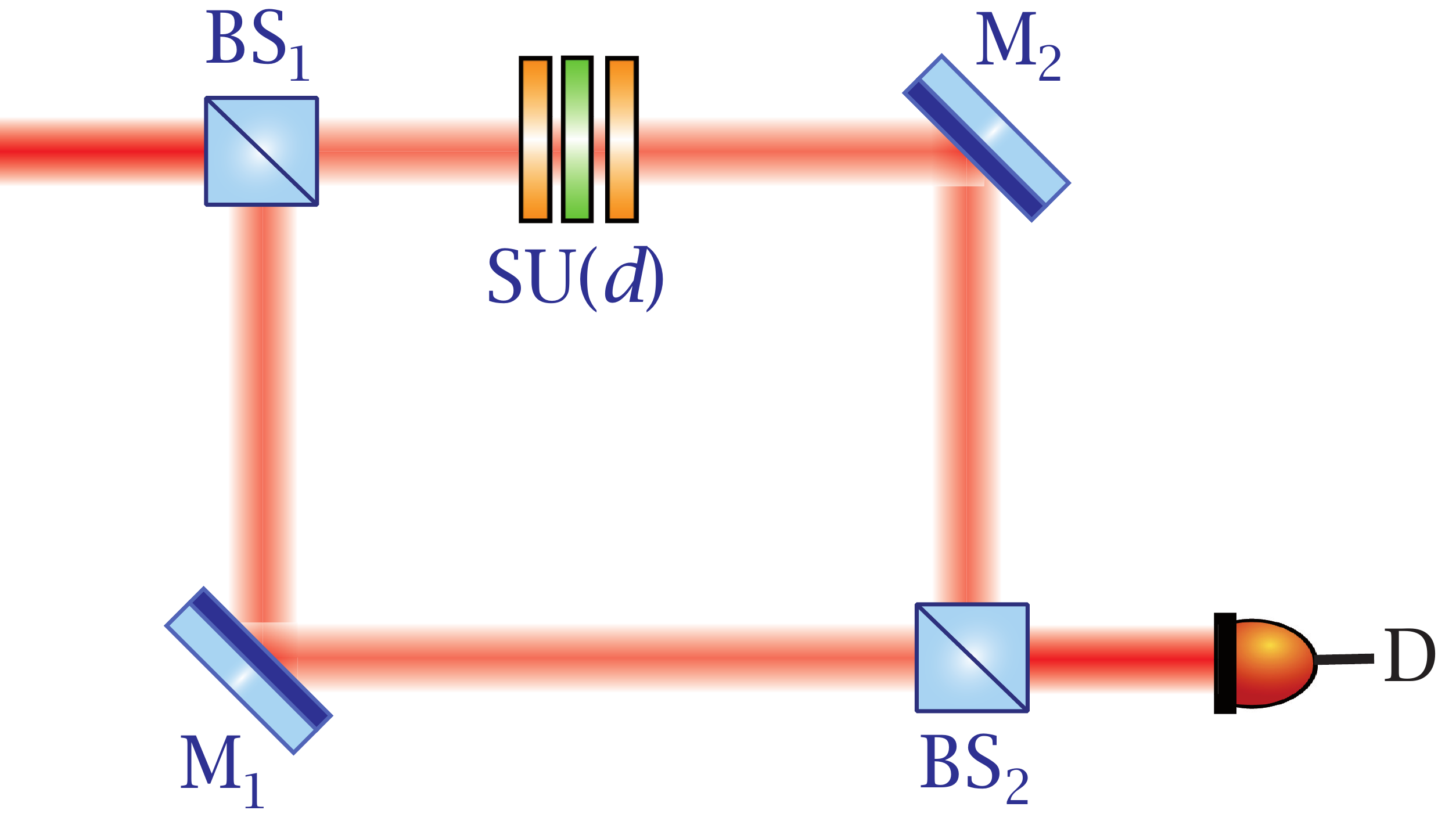}
  \caption{(Color online) Mach-Zehnder setup to interfere the state
    $\varrho$ (lower arm) with its SU($n$) transformed partners (upper
    arm). The distinguishability is related to the visibility of the
    interference pattern at the detector $D$.}
  \label{fig:W_K}
\end{figure}

Consequently, we define a Hilbert-Schmidt degree of polarization as
\begin{equation}
  [ \mathbb{P}_{\mathrm{HS}}^{(n)} ]^{2} = 
  \sup_{g \in \mathrm{SU(n)}}  D^{2}_{\mathrm{HS}} (\rho,\rho_g)
  = \Tr ( \rho^2 ) - 
  \inf_{g \in \mathrm{SU(n)}}  \,  \Tr ( \varrho  \varrho_g ) \, .
\end{equation}
The appealing point is that, formulated in this way,
$\mathbb{P}_{\mathrm{HS}}^{(n)}$ depends on both the purity of the
state and the distinguishability between the state and of 
its rotated counterparts. This later magnitude can be directly
determined as the visibility of an interference experiment, as roughly
schematized in Fig.~1.  As unpolarized states are invariant under
any SU($n$) transformation, this visibility (which is a measure of the
distinguishability between $\varrho$ and $\varrho_{g}$) is zero for
them.

\subsection{Two-dimensional fields}

Let us put the general definition to work for the 2D case. 
The state purity and the distinguishability can be expressed  as
\begin{equation}
  \label{eq:purityBloch}
  \Tr ( \varrho^2 ) = \frac{1}{2}(1+|\mathbf{n}|^2) \, , 
  \quad
  \Tr (\varrho \varrho_{g} ) = \frac{1}{2} 
  (1 +   \mathbf{n}_{\varrho} \cdot \mathbf{n}_{\varrho_{g}} ) \, , 
\end{equation}
where $\mathbf{n}_{\varrho}$ and $\mathbf{n}_{\varrho_{g}}$ are the
Stokes vectors associated with $\varrho$ and $\varrho_{g}$,
respectively.  

To find the minimum overlap, we follow a route that will be useful in
extending this to the 3D case: one notices that any state $\varrho$ can be
brought to a diagonal form $\varrho = R_{g_{0}} \varrho_{0}
R^{\dagger}_{g_{0}}$,  with $R_{g_{0}}$ being an SU(2) matrix and
\begin{equation}
  \label{eq:rhodiag}
  \varrho_{0} = \frac{1}{2}
  \begin{pmatrix}
    1 + | \mathbf{n}|  & 0 \\
    0 & 1- | \mathbf{n}| 
  \end{pmatrix} = 
  \begin{pmatrix}
    \lambda_{+}   & 0 \\
    0 & \lambda_{-} 
  \end{pmatrix} .
\end{equation}
As  $ \inf_{g} \,   \Tr (\varrho \varrho_{g} ) = \inf_{g} \, \Tr
(\varrho_{0} \varrho_{0g} ) $, we get  
\begin{equation}
  \label{eq:min}
  \inf_{g \in \mathrm{SU(2)}}  \Tr (\varrho \varrho_{g} ) =
  \inf_{g \in \mathrm{SU(2)}}   \frac{1}{2} (1 + |\mathbf{n} |^{2}
  \cos \beta  ) =  \frac{1}{2} (1 - |\mathbf{n} |^{2}  )  \, ,
\end{equation}
where $\beta$ is the corresponding Euler angle in
\eqref{eq:su2mat}. The minimum corresponds when $\mathbf{n}_{g}$ is
the antipodal vector $\mathbf{n}_{g} = - \mathbf{n}$, as one might
have anticipated. We thus conclude that
\begin{equation}
  \label{eq:opdefopt}
  \mathbb{P}^{(2)}_{\mathrm{HS}} = 
| \mathbf{n} | = \lambda_{+} - \lambda_{-}  \, ,
\end{equation}
which coincides with the standard definition \eqref{eq:defPeig}.

Notice that in SU(2) we also have
\begin{equation}
  \label{eq:opdefopt2}
  \mathbb{P}^{(2)}_{\mathrm{HS}} =   
  \textstyle{\frac{1}{2}} \displaystyle
  \inf_{g \in \mathrm{SU(2)}}  \Tr
  | \varrho -  \varrho_{g} |  \, ,
\end{equation} 
which shows that the Hilbert-Schmidt $\mathbb{P}^{(2)}_{\mathrm{HS}}$
is proportional to the trace distance. This reinforces the
connection to distinguishability as a measure of the degree of
polarization, since the trace distance is a preferred metric to
quantify the distinguishability between probability
distributions~\cite{Bengtsson:2008ay}.

\subsection{Three-dimensional fields}

Now, we have that 
\begin{equation}
  \label{eq:1}
  \Tr (\varrho^{2}) = \frac{1}{3}(1+2|\mathbf{n}|^2) \, , 
\quad 
 \Tr(\varrho \varrho_{g} ) = 
  \frac{1}{3}(1+2
  \mathbf{n}_{\varrho} \cdot \mathbf{n}_{\varrho_{g}} ) \, .
\end{equation}
This last equation is surprisingly simple, but due to restrictions imposed
by the su(3) algebra, $\mathbf{n}_{\varrho}$ and $\mathbf{n}_{\varrho_{g}}$ cannot be
antiparallel.  Thus, to optimize this distinguishability, we write
again $\varrho = R_{g_{0}} \varrho_{0} R_{g_{0}}^{\dagger}$, with
\begin{eqnarray}
  \label{eq:diagsu3}
  \varrho_{0}
   & = & \frac{1}{3} 
  \left(
    \begin{array}{ccc}
      1 + \sqrt{3} n_3+n_8 & 0 & 0 \\
      0 & 1 - \sqrt{3} n_3+n_8 & 0 \\
      0 & 0 &  1-2 n_8 
    \end{array} 
  \right ) \nonumber \\
& = &  \left (
    \begin{array}{ccc}
      \lambda_{1} & 0 & 0 \\
      0 & \lambda_{2} & 0 \\
      0 & 0 &  \lambda_{3} 
    \end{array} 
  \right )  \, ,
\end{eqnarray}
with the eigenvalues sorted in decreasing order: $\lambda_{1} \ge \lambda_{2}
\ge \lambda_{3}$. The vector $\mathbf{n}_{0}$ associated with $\varrho_{0}$ has two
nonzero components: $n_3$ and $n_8$, and the minimum overlap depends
now on these two parameters.  In addition, positivity imposes 
\begin{eqnarray}
\max \left(-\frac{1 + n_8}{\sqrt{3}}, 
 - \frac{2 - n_8}{\sqrt{3}}\right )
& \le n_3 \le & 
\min \left (\frac{1 + n_{8}}{\sqrt{3}} ,
\frac{2 - n_{8}}{\sqrt{3}} \right ) \, , 
 \nonumber \\
-1 &\le n_8 \le & \frac{1}{2}\, .
\end{eqnarray}
This defines a triangular region of the plane similar to the one
investigated in Ref.~\cite{Saastamoinen:2004lq}. The minimization is now
more involved, and we distinguish two different situations: 
 
\subsubsection{$n_{3}=0$}

This corresponds to a density matrix with two identical eigenvalues:
\begin{equation}
  \varrho_{0}
  =\frac{1}{3} 
  \left(
    \begin{array}{ccc}
      1 + n_8 & 0 & 0 \\
      0 & 1 + n_8 & 0 \\
      0 & 0 &  1-2 n_8 
    \end{array} 
  \right ) \, .
\end{equation}
A direct numerical search shows that the minimum is reached when
$\mathbf{n}_{g}$ is obtained from $\mathbf{n}_{0}$ by the linear transformation 
\begin{equation}
  \left(
    \begin{array}{c}
      n_{3g} \\
      n_{8g} \\
    \end{array}
  \right) =
  \frac{1}{2}
  \left(
    \begin{array}{cc}
      1 & -\sqrt{3} \\
      -\sqrt{3} & -1 \\
    \end{array}
  \right)
  \left(
    \begin{array}{c}
      n_{3} \\
      n_{8} \\
    \end{array}
  \right) \, ,
\end{equation}
so we have $\mathbf{n} \cdot \mathbf{n}_{g} = | \mathbf{n} |^2 \cos(2\pi 
/3) =- n_{8}^{2}/2$. As explained above, the optimal angle between
$\mathbf{n}_{0}$ and $\mathbf{n}_{0g}$ is not $\pi$ because
this angle lies outside the permitted range. That not all angles are
permitted can be explained by the fact that $\varrho_{0}$ and
$\varrho_{0g}$ must have the same eigenvalues since they are unitarily
related. One can confirm that the rotated vector $\mathbf{n}_{g}$
corresponds to the largest eigenvalue being permuted with one of the
smaller. Hence,  we can recast the infimum as Tr$(\varrho \varrho_g)=
2\lambda_1 \lambda_3+\lambda_3^2 $, so the  degree becomes
\begin{equation}
\label{eq:lho}
\mathbb{P}^{(3)}_{\mathrm{HS}} = \lambda_{+} - \lambda_{-} \, ,
\end{equation}
where $\lambda_{+} = \lambda_{1}$ and $\lambda_{-} = \lambda_{3} $
(here, $\lambda_{2} =\lambda_{3}$). In this way, it appears as the
natural generalization of the 2D version \eqref{eq:defPeig}.

\subsubsection{$n_{3} \neq 0$}
 The three eigenvalues are now different. We set $n_8=X\,n_3$ 
and write
\begin{equation}
  \left ( 
    \begin{array}{cc}
      n_{3g}\\
      n_{8g}
    \end{array}
  \right)
  =
  \left (
    \begin{array}{cc}
      \cos\theta& \sin\theta\\
      \sin\theta&\cos\theta
    \end{array}
  \right ) 
  \left (
    \begin{array}{cc}
      -1&0\\
      0&1
    \end{array}
  \right )
  \left (
    \begin{array}{cc}
      n_{3}\\
      n_{8}
    \end{array}
  \right) \, . 
  \label{overlaptransformation}
\end{equation}

We have to consider three different zones:
\begin{itemize}
\item[1.] $X>1/\sqrt{3}$.   
  The minimum is found when the angles $\beta_1$, $\beta_2$, and
  $\beta_3$ are respectively set to $( \pi,\pi, \pi )$ in the SU(3)
  matrix \eqref{su3} and the rest of the angles equal 0.  Then, 
$\theta=2\pi/3$ in \eqref{overlaptransformation} reproduces
  this minimum.

\item[2.] $| X | < 1/\sqrt{3}$.  
The minimum is now found when the angles $\beta_1,\beta_2$ 
and $\beta_3$ take the values $(0, \pi, 0)$ in \eqref{su3}.  Here,
$\theta=0$ in  \eqref{overlaptransformation} gives  the correct result. 

\item[3.] $X<-1/\sqrt{3}$.  
  Here, the minimum occurs for $(\beta_1,\beta_2,\beta_3)=(0, 0,
  \pi)$, corresponding to the angle $\theta=-2\pi/3$ in
  \eqref{overlaptransformation}.
\end{itemize}
The transformed density matrix accounts for a reshuffling of the
eigenvalues and by simple inspection one can check
that \eqref{eq:lho}  holds for all three cases.

\begin{figure}
\centering{
\includegraphics[width=0.97\columnwidth]{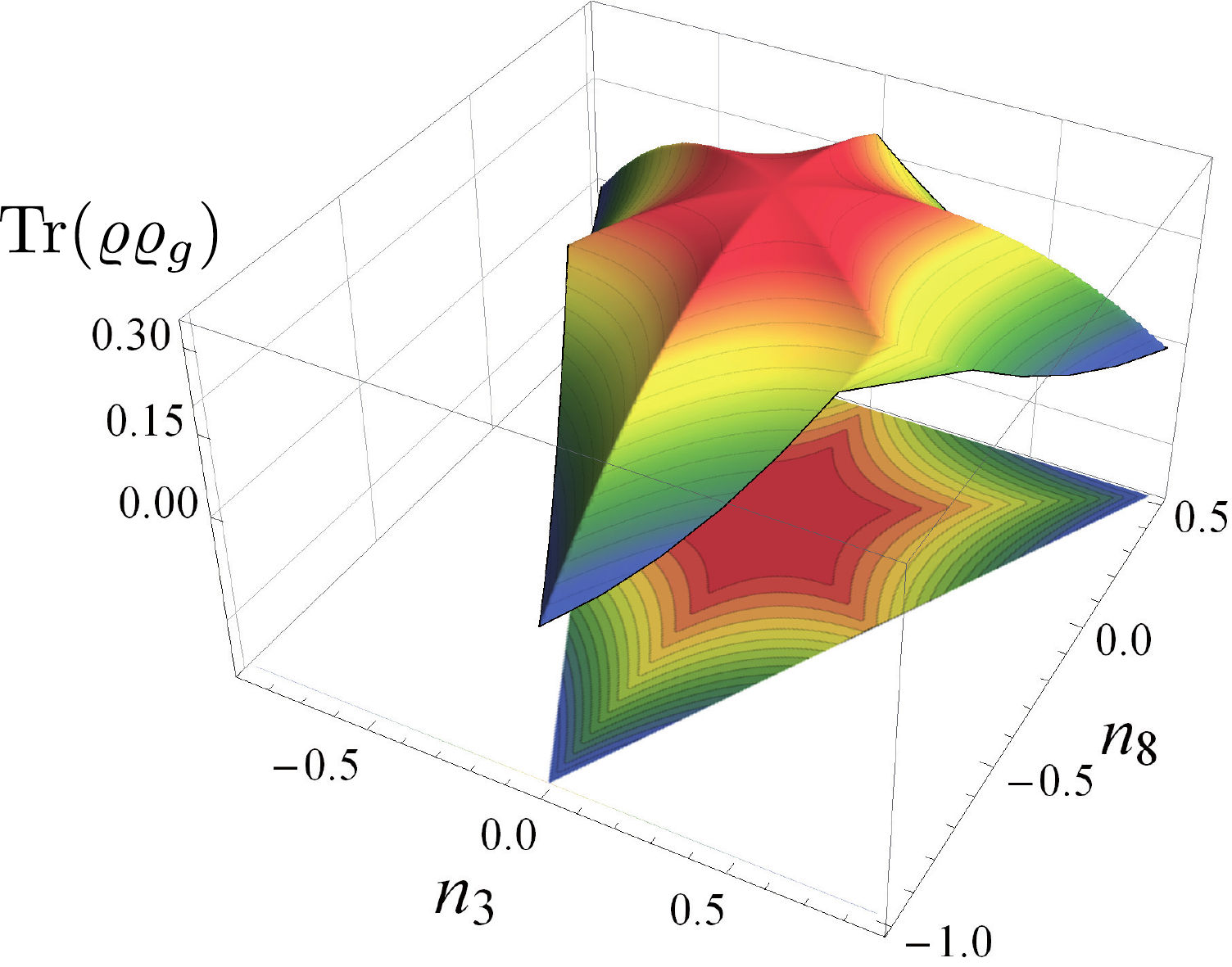}}
\caption{(Color online) A 3D plot locating the overlap $\Tr (\varrho
  \varrho_{g})$ as a function of the parameters $n_3$ and $n_8$. At
  the bottom, we show a contour plot of the surface.}
  \label{fig:starfish}
\end{figure}

Additional insight can be gained by considering a
three-dimensional plot illustrating the loci of the minima and a
contour plot of these points, as shown in Fig.~\ref{fig:starfish}.
The 6-fold symmetry of the result (corresponding to the six possible
permutations of $\lambda_1,\lambda_2,\lambda_3$, so they remain in
decreasing order) is explicit and quite similar to the symmetry
exploited in Ref.~\cite{Sheppard:2011hc}.

The Hilbert-Schmidt degree (\ref{eq:opdefopt2})  admits a direct 3D
translation, namely, 
\begin{equation}
  \label{eq:opdefopt3}
  \mathbb{P}^{(3)}_{\mathrm{HS}} =
  \textstyle{\frac{1}{2}} \displaystyle
  \inf_{g \in \mathrm{SU(3)}}  \Tr
  | \varrho -  \varrho_{g} |  \, .
\end{equation}

In this respect, it is worth stressing that several 3D measures have
already been  introduced in terms of the eigenvalues of the $3\times 3$
polarization matrix. Relevant examples are~\cite{Sheppard:2011hc}
\begin{equation}
  \label{eq:6}
  \mathbb{P}_{\mathrm{PP}}^{(3)} = \lambda_{1} - \lambda_{2} \, ,
  \qquad 
  \mathbb{P}_{\mathrm{U}}^{(3)} = 3 \lambda_{3} 
  \qquad
  \mathbb{P}_{\mathrm{PU}}^{(3)} = 2 (\lambda_{2} -  \lambda_{3} ) \,. 
\end{equation}
Here, $\mathbb{P}_{\mathrm{PP}}^{(3)}$ measures the strength of the pure
polarized component, $\mathbb{P}_{\mathrm{U}}^{(3)}$ is the strength of the
unpolarized component, and $\mathbb{P}_{\mathrm{PU}}^{(3)}$ is the strength
of the component that is unpolarized within a plane. In
Ref.~\cite{Gamel:2014uf}, the method of majorization, previously used
in quantum information, is applied to these measures to establish a
partial ordering on the polarization state spaces. 

\begin{figure}
  \begin{center}
    \noindent
    \includegraphics[width=0.45\linewidth]{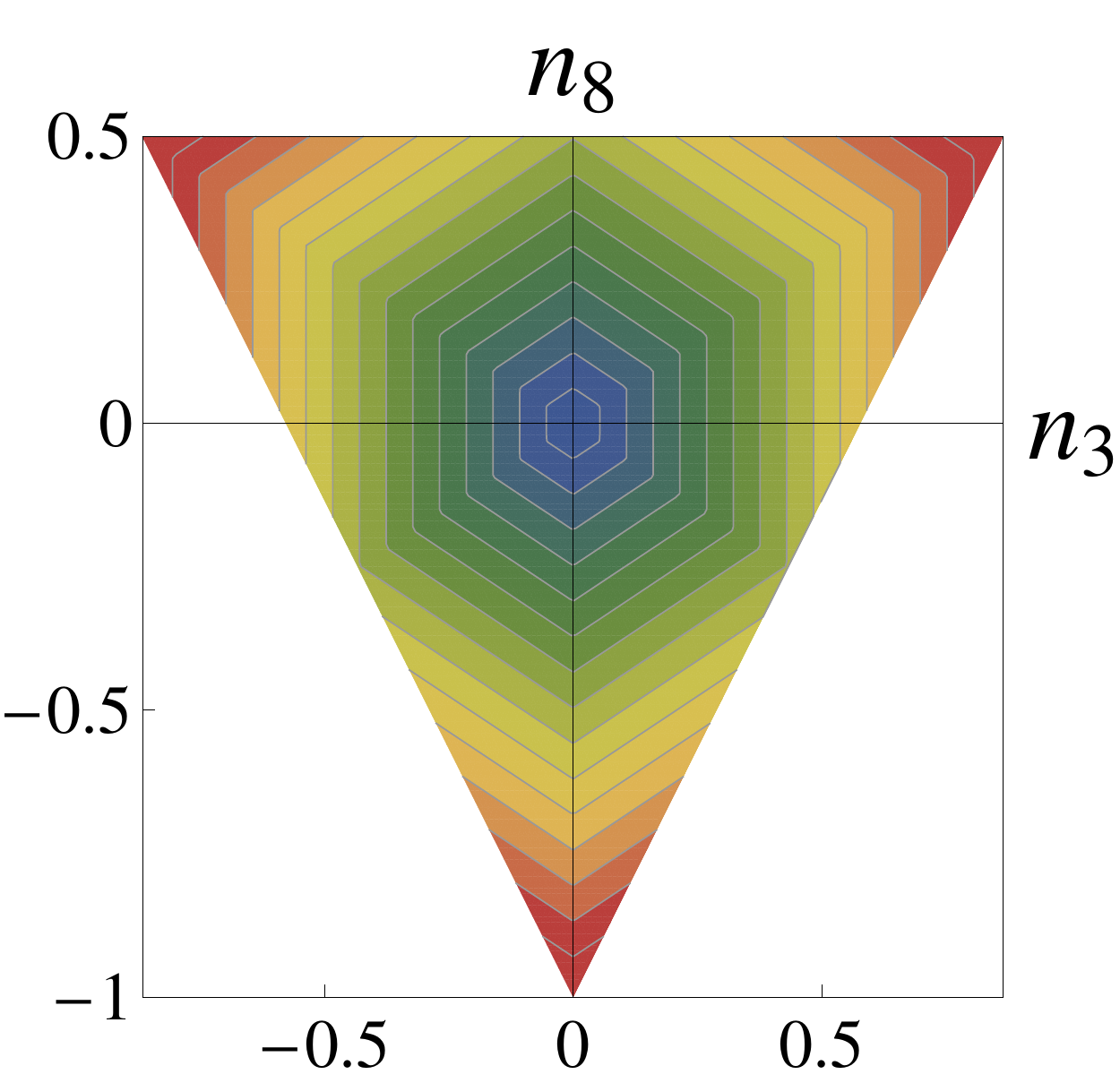}
    \hfill
    \includegraphics[width=0.45\linewidth]{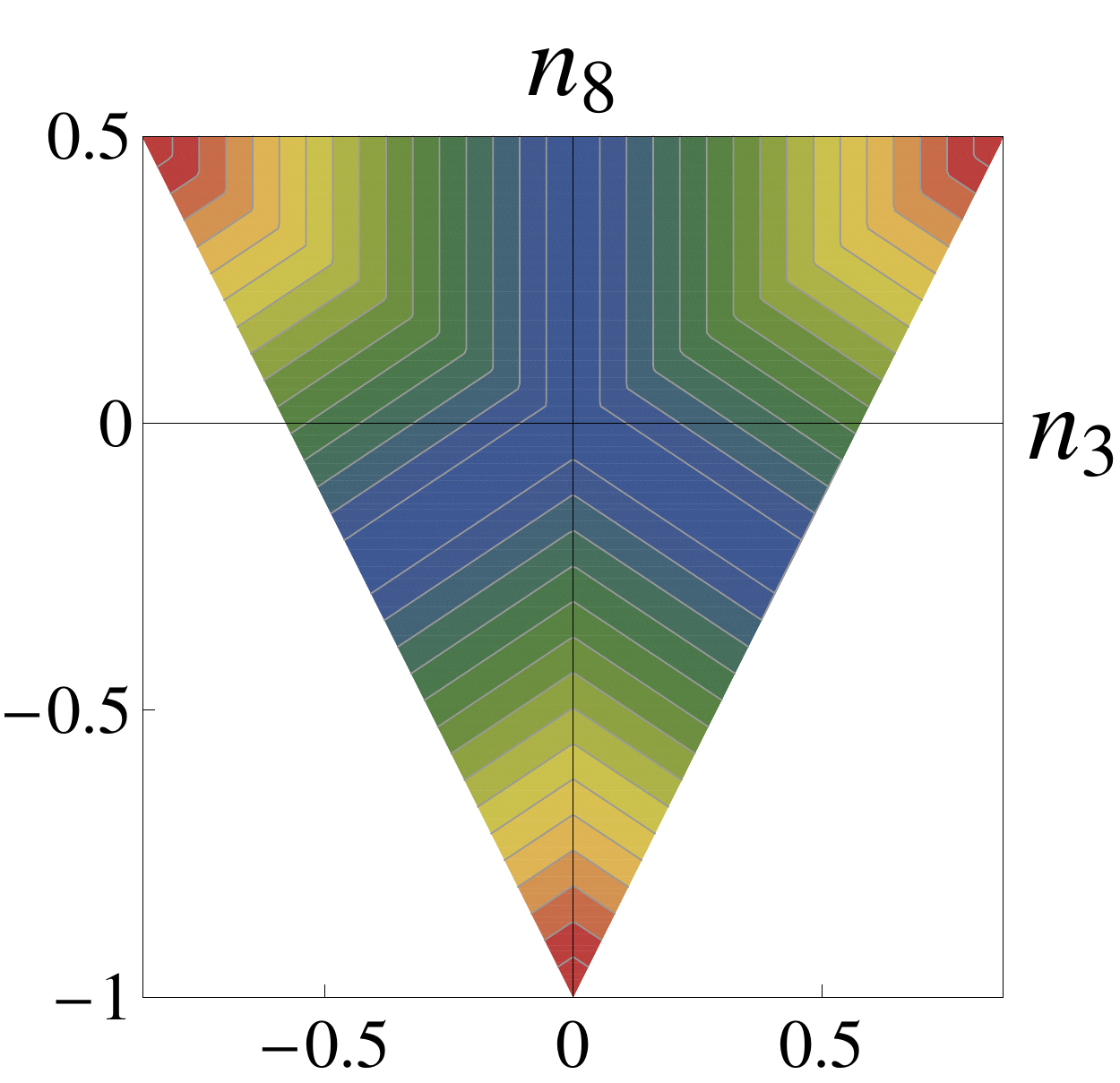}\\
    \includegraphics[width=0.45\linewidth]{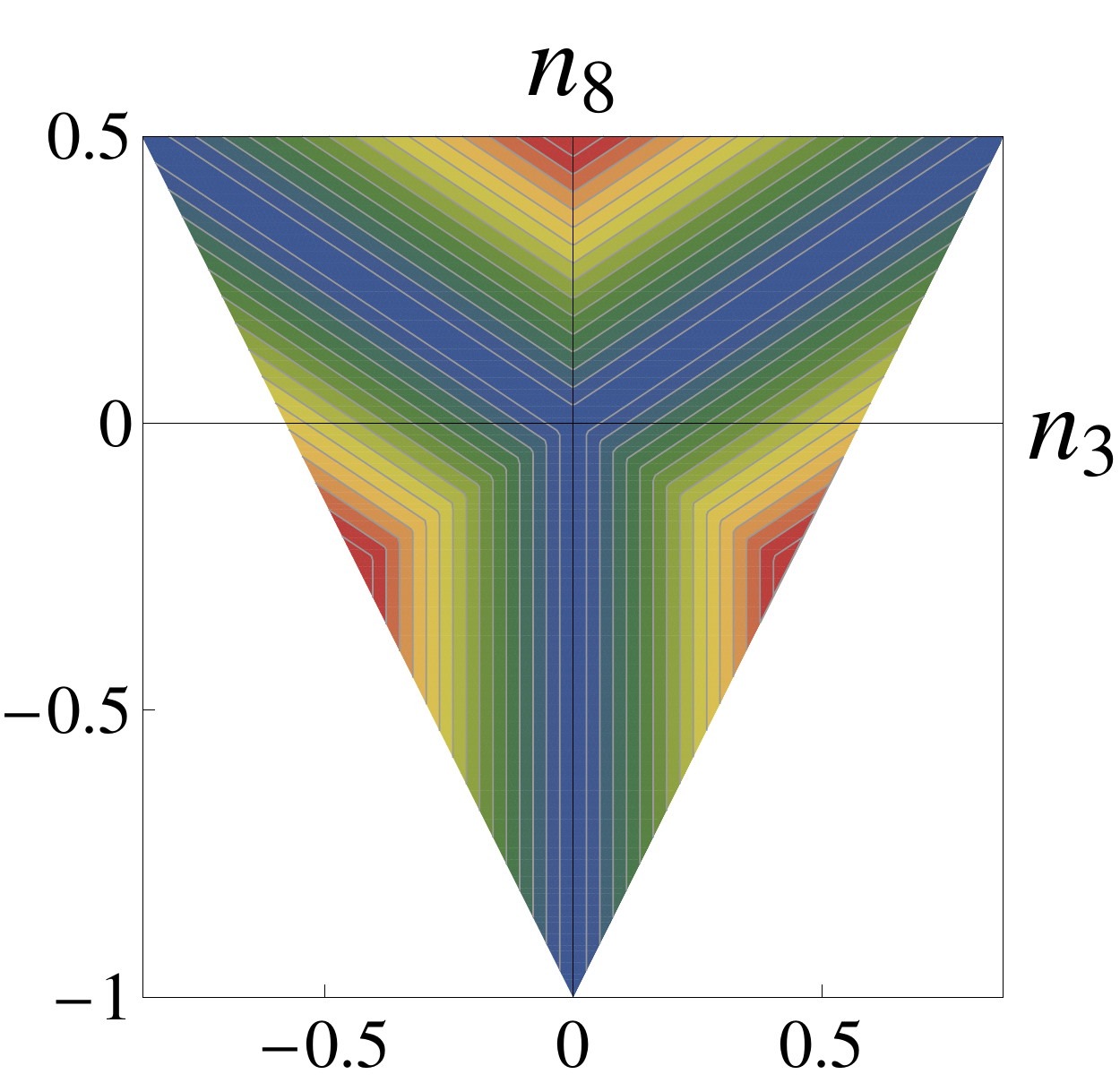}
    \hfill
    \includegraphics[width=0.45\linewidth]{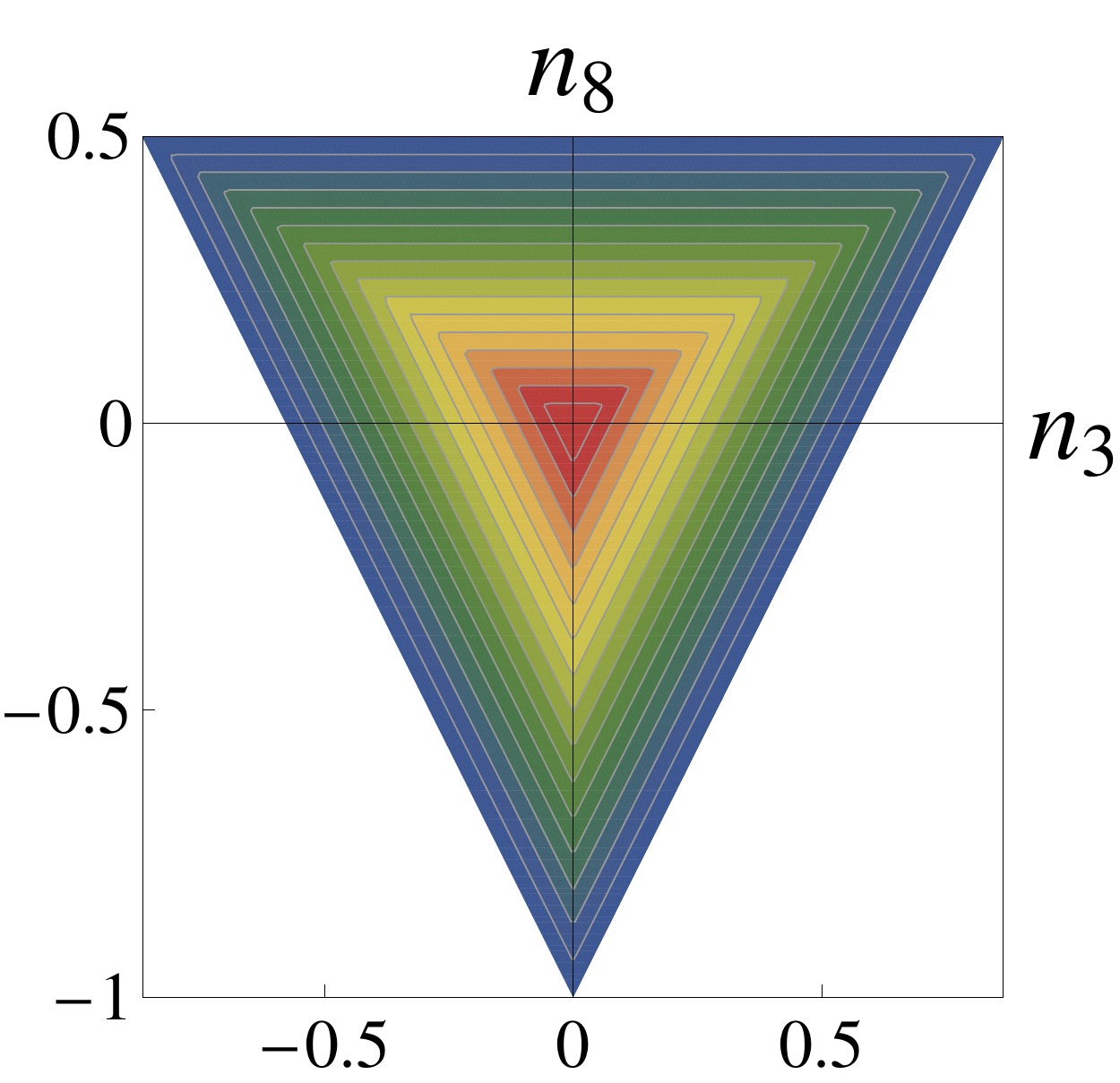}
    \caption{(Color online) Isocontour lines (in the same color scale
      as in Fig.~2) of the different degrees of polarization as a
      function of the parameters $n_3$ and $n_8$. In the top panel,
      $\mathbb{P}_{\mathrm{HS}}^{(3)}$ (left) and
      $\mathbb{P}_{\mathrm{PP}}^{(3)}$ (right); in the bottom panel,
      $\mathbb{P}_{\mathrm{PU}}^{(3)}$ (left) and
      $\mathbb{P}_{\mathrm{U}}^{(3)}$ (right).}
    \label{fig:heralded}
  \end{center}
\end{figure}

For the sake of completeness, in Fig.~\ref{fig:heralded} we have
plotted the lines of constant degrees of polarization for
$\mathbb{P}_{\mathrm{HS}}^{(3)}$ and the three alternatives in
\eqref{eq:6}, again as a function of $n_{3}$ and $n_{8}$. The figure
is so explicit that it does not deserve many additional comments. What is
really remarkable is how differently these measures quantify the
polarization at the apices of the triangle.

\section{Concluding remarks}
\label{sec:conc}

We have explored the use of a degree of polarization based on the
distance of a state to the set of its rotated counterparts. Such a
definition is closely related to other recent proposals in
different areas of quantum optics and is well behaved in the classical
domain, providing an operational approach that can be extended from
the 2D formalism (where it reproduces the standard results) to the 3D
case (where it gives a new measure). 

The resulting degree is tightly linked to the notion of
distinguishability, which can be experimentally determined as the
visibility in a simple interference setup, which confirms previous
contentions along the same lines~\cite{Klyshko:1992wd}.

We hope that our analysis adds to and clarifies the discussion on measures
of higher-dimensional polarization in the literature.

\begin{acknowledgments}
  The work of G. B. is supported by the Swedish Foundation for
  International Cooperation in Research and Higher Education (STINT)
  and the Swedish Research Council (VR) through its Linn{\ae}us Center
  of Excellence ADOPT and contract No.~621-2011-4575. H. d G.  is
  supported by the Natural Sciences and Engineering Research Council
  (NSERC) of Canada. A. K. is thankful for the financial assistance of
  the Mexican CONACyT (Grant No.~106525).  Finally, P. H. and
  L. L. S. S. acknowledge the support from the Spanish MINECO (Grant
  FIS2011-26786). It is also a pleasure to thank I.~Bengtsson,
  J.~J.~Monz\'on, and G.~Leuchs for stimulating discussions.
\end{acknowledgments}

\appendix

\section{Basic facts and parametrization of SU(3)}

The su(3) algebra is usually presented in terms of a set of Hermitian
generators known as the Gell-Mann matrices~\cite{Weigert:1997fk}
${\Lambda}_r$ ($r= 1, \ldots, 8$). They obey the commutation relations
\begin{equation} 
 [ {\Lambda}_r, {\Lambda}_s ] = 2 i f_{rst}  {\Lambda}_t ,
\end{equation}
where, above and in the following, the summation over repeated indices
applies. The structure constants $f_{rst}$ are elements of a
completely antisymmetric tensor spelled out explicitly in
Ref.~\cite{Arvind:1997qf}, whose notation we follow.

A particular feature of the generators of SU(3) in the defining $3
\times 3$ matrix representation is closure under
anticommutation
\begin{equation}
  \{ {\Lambda}_r, {\Lambda}_s \} =
  \frac{4}{3} \delta_{rs} {\openone}
  + 2  d_{rst} {\Lambda}_t ,
\end{equation}
where $\delta_{rs}$ is the Kronecker symbol and $d_{rst}$ form a
totally symmetric tensor~\cite{Weigert:1997fk}. 

For the following, a vector-type notation is useful, based on the
structure constants. The $f$ and $d$ symbols allow us to define both
antisymmetric and symmetric products by
\begin{eqnarray}
  (\mathbf{A} \wedge \mathbf{B})_r & = &
  f_{rst} A_s B_t =
  -(\mathbf{B} \wedge \mathbf{A})_r  \, , \nonumber \\
  & & \\
  (\mathbf{A} \star \mathbf{B})_r & = &
  \sqrt{3} d_{rst} A_s B_t =
  + ( \mathbf{B} \star \mathbf{A} )_r  \, .
  \nonumber
\end{eqnarray}

Given a density matrix ${\varrho}$ we can expand it in terms of the
unit matrix $\openone$ and the ${\Lambda}_{r}$ in the form
\begin{equation} 
\rho  = \frac{1}{3}(1+ \sqrt{3} \, \mathbf{n} \cdot
  {\bm{\Lambda}} ) .
\end{equation}
This is the equivalent to the Bloch ball for SU(3). For a pure
state the analogous Bloch sphere is defined by the condition
\begin{equation}
  \mathbf{n} \cdot \mathbf{n}=1,
  \qquad
  \mathbf{n} \star \mathbf{n}= \mathbf{n} .
\end{equation}
Thus, each pure qutrit state corresponds to a unique unit vector
$\mathbf{n} \in \mathcal{S}^7$, the seven-dimensional unit sphere. In
addition, this vector must obey the condition $\mathbf{n} \star
\mathbf{n} = \mathbf{n}$, which places three additional constraints,
thus reducing the number of real parameters required to specify a pure
state from seven to four.


%

\end{document}